\documentclass[aps,prd,preprint,tightenlines,groupedaddress,showpacs]{revtex4}
\usepackage{epsf,epsfig,graphics,graphicx}
\begin{document}

\title{Quantum noise and large-scale cosmic microwave background anisotropy}

\author{Chun-Hsien Wu$^1$}
\author{Kin-Wang Ng$^1$}
\author{Wolung Lee$^2$}
\author{Da-Shin Lee$^3$}
\email{dslee@mail.ndhu.edu.tw}
\author{Yeo-Yie Charng$^1$}
\affiliation{
$^1$Institute of Physics, Academia Sinica, Taipei, Taiwan 115, R.O.C.\\
$^2$Department of Physics, National Taiwan Normal University,
Taipei, Taiwan 116, R.O.C.\\
$^3$Department of Physics, National Dong Hwa University,
Hua-Lien, Taiwan 974, R.O.C.}

\date{\today}

\begin{abstract}
We propose a new source for the cosmological density perturbation
which is passive fluctuations of the inflaton driven dynamically by
a colored quantum noise as a result of its coupling to other massive
quantum fields. The created fluctuations grow with time
during inflation before horizon-crossing. However, the larger-scale
modes cross out the horizon earlier, thus resulting in a suppression
of their density perturbation as compared with those on small
scales. By using current observed CMB data to constrain the parameters
introudced, we find that a significant contribution from the
noise-driven perturbation to the density perturbation is still
allowed.  It in turn gives rise to a suppression of the large-scale
CMB anisotropy that may be relevant to the observed low quadrupole
in the WMAP CMB anisotropy data. We also briefly discuss the
implications to the energy scale of inflation and the spectral index and 
non-Gaussianity of the density perturbation.
\end{abstract}

\maketitle

\section{Introduction}

The inflationary scenario~\cite{olive}, in which the present Universe
is only a small local patch of a causally connected region at early time
which underwent an exponential expansion driven by the inflaton potential,
is generally accepted for explaining the observed spatially
flat and homogeneous Universe. In addition,
its quantum fluctuations during inflation give rise to primordial
Gaussian density fluctuations with a nearly scale-invariant power
spectrum, which is consistent with recent astrophysical and cosmological
observations such as structure formation and CMB anisotropies~\cite{texas04}.

Despite its success, some basic problems still remain unsolved.
What is the origin of the inflaton potential?
Do classical density imhomogenities that
we observe today really come from quantum
fluctuations of the inflaton?
To answer these questions is a big challenge to both
theories and observations. The CMB anisotropy measurements made by
COBE~\cite{cobe} and recently confirmed by WMAP~\cite{wmap} have
shown that the amplitudes of the low-$l$ multipoles are lower than
expected for the $\Lambda$CDM model, although the statistical
significance of such an anomaly is not large~\cite{efsta}. If this
is true, it will give a clue to probe the physics of inflation.

There are many proposed solutions for the anomaly, mostly based on
some {\it ad hoc} new ingredients in the generating process of
density perturbation~\cite{fang}. Recently the colored noise has
been considered to explain the anomaly in the context of stochastic
inflation~\cite{mata}. In the stochastic approach to
inflation~\cite{star}, the inflaton field is coarse-grained by a
Heaviside window function with a sharp frequency cutoff. Thus, the
high-frequency modes constitute a white noise which drives the
dynamics of the coarse-grained inflaton and generates a
scale-invariant density power spectrum. In Ref.~\cite{mata}, they
have adopted instead a Gaussian window function with a width
characterizing the size of the coarse-grained domain~\cite{wini}
which is then arbitrarily chosen to be comparable to the Hubble
radius. This smooth window function results in a colored noise which
generates fewer fluctuations than the white noise on the scales
slightly less than the chosen coarse-grained domain. So, they find a
blue tilt of the power spectrum on large scales which can be tuned
to fit the WMAP large-scale anisotropy data. Nevertheless, in this
approach they inevitably resort to an {\it ad hoc}
smoothing window function and therefore the width of the window
function remains undetermined.

Apart from this free-field treatment, however, the inflaton is
expected to be interacting with other fields; otherwise, the
potential energy stored in the inflaton field cannot be converted
into radiation after inflation. Cosmological phenomena associated
with an interacting inflaton have been immensely studied in the
context of reheating, preheating, backreactions to the inflaton
dynamics, and etc.~\cite{linde}. In particular, the influence
from the interaction with other fields on density fluctuations
during inflation has been discussed~\cite{noise}. These have not
only provided us with a field theoretical framework to understand
the underlying physics of the inflation, but also explored the
possibility of generating primordial density fluctuations with
statistical properties deviated from both the scale-invariance and
the Gaussianity.

In this paper, we will study the possible effects of coupling massive
quantum fields to inflation and constrain the couplings and mass
parameters by using observed CMB data.  We will show that the
inflaton fluctuations driven by the colored noise due to the
interaction with the other field are strongly dependent on the onset
of inflation and become scale-invariant asymptotically at small
scales. This is due to the fact that the earlier a Fourier mode of
the noise leaves the horizon the shorter the time can it drive the
growth of the corresponding Fourier mode of the inflaton
fluctuations. However, these passive fluctuations would not grow
arbitrarily with time but only get boosted in an intrinsic time
scale. This would result in a suppression of the density power
spectrum on large scales which may give a theoretical support to the
low CMB low-$l$ multipoles, if the passive ones contribute a
significant portion to the density perturbation.

\section{Influence functional approach}

We will adopt the influence functional method~\cite{fey} to take
into account the quantum fluctuation effects from the massive fields coupled
to the inflaton in a real-time manner, instead of using an one-loop 
effective potential as usually found for example in Ref.~\cite{ala}.   
Thus, the effective Langevin equation of the inflaton is obtained,
describing the time dependent corrections of the slow-rolling dynamics
of the inflaton originally given by the classical inflaton potential.
In particular, this Langevin equation, which goes beyond the mean field 
approximation, involves a stochastic noise term that may drive the growth 
of perturbation of the inflaton as we will see later.

\subsection{Our model and Langevin equation for inflaton}

Let us consider a slow-rolling inflaton $\phi$ coupled to a massive
scalar field $\sigma$ with a Lagrangian given by
\begin{equation}
\mathcal{L}=\frac{1}{2}g^{\mu\nu}\partial_{\mu}\phi\,\partial_{\nu}\phi+
            \frac{1}{2}g^{\mu\nu}\partial_{\mu}\sigma\,\partial_{\nu}\sigma
             -V(\phi)-\frac{m_{\sigma}^{2}}{2}\sigma^{2}-
             \frac{g^{2}}{2}\phi^{2}\sigma^{2},
\label{model}
\end{equation}
where $V(\phi)$ is the inflaton potential that complies with the
slow-roll conditions and $g$ is a coupling constant. Thus, we can
approximate the space-time during inflation by a de Sitter metric
given by
\begin{equation}
ds^{2}=a^2 (\eta)(d\eta^{2}-d{\bf x}^2),
\end{equation}
where $\eta$ is the conformal time and $a(\eta)= -1/(H\eta)$ with
$H$ being the Hubble parameter. Here we rescale $a=1$ at the initial
time of the inflation era, $\eta_i= -1/H$. As for the
system-environment field splitting, one may separate the inflaton
field into its low- and high-frequency modes by introducing a window
function.  The low-frequency modes of cosmological relevance can be
regarded as the system of interest whereas the high-frequency 
inflaton modes as well as the field $\sigma$ are treated as an
environment being traced out to the extent in how they influence the
system. Nevertheless, the perturbation of the inflaton of
cosmological relevant scales are found to be insignificantly
affected from its high-frequency counterparts due to inflaton
self-interaction~\cite{noise}. Thus, here we restrict ourselves to
the influence of quantum fluctuations from the field $\sigma$ on
the inflaton perturbation. In addition, the coupling of the inflaton 
to $\sigma$ results in renormalization of the $\sigma$ mass term due to
quantum fluctuations of the inflaton, which will be included in the
later discussion.

To proceed, let us assume that the initial density matrix at time
$\eta_i$ can be factorized as
\begin{equation}\label{initialcond}
    \rho(\eta_i)=\rho_{\phi}(\eta_i)\otimes\rho_{\sigma }(\eta_i)\,.
\end{equation}
The full density matrix evolves unitarily and the evolution 
can be described by employing the closed-time-path
formalism. Following the influence functional
approach~\cite{fey,noise}, we trace  out the field $\sigma$ in the
perturbative expansion. The reduced density matrix of the system
then becomes
\begin{equation}
\rho_{r}(\phi_{f},\phi'_{f};\eta_{f})=\int d\phi_{i} \, d\phi'_{i}
\, \mathcal{J} (\phi_{f}, \phi'_{f}, \eta_{f} ; \phi_{i}, \phi'_{i},
\eta_{i} ) \,\rho_{r}(\phi_{i},\phi'_{i};\eta_{i}) \, ,
\end{equation}
where the propagating function $\mathcal{J} (\phi_{f}, \phi'_{f},
\eta_{f} ; \phi_{i}, \phi'_{i}, \eta_{i} )$ is obtained as
\begin{equation}
\mathcal{J} (\phi_{f}, \phi'_{f}, \eta_{f} ; \phi_{i}, \phi'_{i},
\eta_{i} )=\int_{\phi_{i}}^{\phi_{f}} \, \mathcal{D}\phi^+
 \int_{\phi'_{i}}^{\phi'_{f}} \, \mathcal{D}\phi^- \,
e^{i\left( S_0 [\phi^+]-S_0 [\phi^-] \right)} \times 
e^{i S_{IF}[\phi^+,\phi^-]} \label{propagatingfun}
\end{equation}
and the action for the field $\phi$ is given by
\begin{equation}
S_0 [ \phi ]= \int  d^4 x \, a^2 (\eta) \left[ \, \frac{1}{2}
\left(\frac{d\phi}{d\eta}\right)^2 -\frac{1}{2} \left( \nabla \phi \right)^2
             - a^2(\eta) \, V(\phi) \right] \, .
\end{equation}
Here we obtain the influence functional up to order $ g^4$ as
\begin{eqnarray}
e^{i S_{IF}[\phi^+,\phi^-]}   & = & \exp \left\{ -i
\frac{g^2}{2} \int  d^{4} x_{1}  \, a^4 (\eta_1) \, \left[ \phi^{+
2}(x_{1}) \, \langle \sigma^+ (x_1) \sigma^+ (x_1) \rangle - \phi^{-
2}(x_{1}) \, \langle
\sigma^- (x_1) \sigma^- (x_1) \rangle \right] \right. \nonumber \\
& &  - \,\frac{g^{4}}{4} \int d^{4} x_{1}  \int
d^{4} x_{2}   \, a^4 (\eta_1) \, a^4(\eta_2) \nonumber \\
&&  \left[ \phi^{+ 2}(x_{1}) \, \langle \sigma^+ (x_1) \sigma^+
(x_2) \rangle^2 \, \phi^{+2}(x_{2}) -  \phi^{+ 2}(x_{1}) \, \langle
\sigma^+ (x_1)
\sigma^- (x_2) \rangle^2 \, \phi^{-2}(x_{2}) \right. \nonumber \\
 &  &  \left. \left.-  \phi^{- 2}(x_{1}) \, \langle \sigma^- (x_1) \sigma^+ (x_2)
\rangle^2 \, \phi^{+2}(x_{2}) +  \phi^{- 2}(x_{1}) \, \langle
\sigma^- (x_1) \sigma^- (x_2) \rangle^2 \, \phi^{-2}(x_{2}) \right]
\right\} \, . \label{influencefun}
\end{eqnarray}
The Green's functions of the $\sigma$ field are defined by
\begin{eqnarray}
    \bigl< \sigma^+ (x) \sigma^+ (x')\bigr>&=& \bigl< \sigma (x) \sigma(x')\bigr> \,\theta(\eta-\eta')
    +\bigl< \sigma(x')\sigma(x) \bigr> \,\theta (\eta'-\eta)\,,\nonumber\\
    \bigl< \sigma^- (x) \sigma^- (x')\bigr> &=& \bigl< \sigma (x') \sigma(x)\bigr>\,\theta(\eta-\eta')
    +\bigl< \sigma(x)\sigma(x')\bigr> \,\theta (\eta'-\eta)\,,\nonumber\\
    \bigl< \sigma^+ (x) \sigma^- (x')\bigr>&=&\bigl< \sigma (x) \sigma(x')\bigr>\,,\nonumber\\
    \bigl< \sigma^- (x) \sigma^+ (x')\bigr>&=& \bigl< \sigma(x')\sigma(x)\bigr> \,,
\end{eqnarray}
and can be explicitly constructed as long as its vacuum state has
been specified. To obtain the semiclassical Langevin equation, it is
more convenient to introduce the average and relative field
variables:
\begin{equation}
 \phi =\frac{1}{2}( \phi^+ + \phi^- ) \,\,\, , \,\,\,   
 \phi_\Delta=\phi^+ - \phi^- \, .
\end{equation}
The coarse-grained effective action(CGEA) including the influence 
action $S_{IF}$ obtained from
Eqs.~(\ref{propagatingfun}) and~(\ref{influencefun}) is then given by
\begin{eqnarray}
S_{CGEA} \left[ \phi, \phi_{\Delta} \right]&=& \int d^{4} x \,
a^2 (\eta) \, \phi_{\Delta} (x) \left\{ -\ddot{\phi}(x) -
2aH\dot{\phi}(x) +\nabla^{2}\phi(x)  - a^{2}\left[ V'(\phi)
+g^{2}\langle\sigma^{2}\rangle\phi(x) \right] \right. \nonumber \\
&& \left.
\,\,\,\,\,\,\,\,\,\,\,\,\,\,\,\,\,\,\,\,\,\,\,\,\,\,\,\,\,\,\,\, -
g^4 a^2 (\eta) \phi (x)  \int d^{4} x' \, a^4(\eta')
\,\theta(\eta-\eta')\, i G_- (x,x') \phi^2 (x') \right\}\,
\nonumber \\
&+&  i \frac{g^4}{2} \int  d^{4} x  \int d^{4} x' a^4 (\eta) a^4
(\eta') \, \phi_{\Delta} (x) \phi (x) \, G_+ ( x,x') \,
\phi_{\Delta} (x') \phi (x') + \mathcal{O} ( \phi_{\Delta}^3) \, ,
\nonumber \\
\end{eqnarray}
where the dot and prime denote respectively differentiation with
respect to $\eta$ and $\phi$. In addition, we have used the fact
that correlation functions of the fields evaluated at the same
space-time point in the $+$ and $-$ branches are equal, namely,
$\langle \sigma^+ (x_1) \sigma^+ (x_1) \rangle=\langle \sigma^-
(x_1) \sigma^- (x_1) \rangle \equiv \langle \sigma^2 (x_1) \rangle$.
The kernels $G_{\pm}$ can be obtained from the Green's function of
$\sigma$:
\begin{equation}
G_{\pm}(x,x')=\langle\sigma(x)\sigma(x')\rangle^2 \pm
                               \langle\sigma(x')\sigma(x)\rangle^2.
\label{grfct}
\end{equation}
The imaginary part of the above influence action can be
re-expressed by introducing an auxiliary field $ \xi $  with a
distribution function of the Gaussian form,
\begin{equation}
P[\xi ] = \exp \left\{ - \frac{1}{2}  \int d^{4}x \, \int d^{4} x'
\, \xi (x) \, G_{+}^{-1} (x,x') \, \xi (x') \right\} \, ,
\label{noisedistri}
\end{equation}
leading to
\begin{equation}
e^{i S_{CGEA}}  =  \int {\cal D} \xi \, P [\xi ] \, \exp i S_{\rm
eff} \left[ \phi , \phi_{\Delta}, \xi \right] \, ,
\end{equation}
with the effective action $ S_{\rm eff}$ given by
\begin{eqnarray}
S_{\rm eff} [\phi,\phi_{\Delta}, \xi ] && = \int d^4 x \, a^2 (\eta)
\, \phi_{\Delta} (x) \left\{ -\ddot{\phi}(x) - 2aH\dot{\phi}(x)
+\nabla^{2}\phi(x)  - a^{2}\left[ V'(\phi)
+g^{2}\langle\sigma^{2}\rangle\phi(x) \right] \right. \nonumber \\
&&-\left.   g^4 a^2 (\eta) \phi (x) \int d^{4} x' \, a^4(\eta') \,
\theta(\eta-\eta') \, i G_- (x,x') \phi^2 (x')    + g^2 a^2 (\eta)
\phi (x) \xi(x) \right\} \, .
\end{eqnarray}
The semiclassical approximation requires to extremize the effective
action $ \delta S_{\rm eff}/ \delta \phi_{\Delta}$ when long-wavelength 
inflaton modes of cosmological interest have gone through the 
quantum-to-classical transition due to the rapid expansion of the scale
factor~\cite{sta}. Then, we obtain the semiclassical Langevin
equation for $\phi$:
\begin{eqnarray}
&& \ddot{\phi}+2aH\dot{\phi}-\nabla^{2}\phi+a^{2}\left[V'(\phi)
+g^{2}\langle\sigma^{2}\rangle\phi\right]-g^{4}a^{2}{\phi}  \int
d^{4} x'
 a^4(\eta') \times \nonumber \\
&& \,\,\,\,\,\,\,\,\,\,\,\,\,\,\,\,\,\,\,\,\,
\theta(\eta-\eta')\,i\, G_{-}(x,x') {\phi}^{2}(x')=g^2 a^2 \, \phi
\, \xi+\xi_w \, , \label{lange}
\end{eqnarray}
where we have included the white noise in the free-field stochastic
inflation~\cite{star}. The white noise $\xi_w$ reproduces the active
or intrinsic inflaton quantum fluctuations
$\langle\varphi_q^{2}\rangle$ with a scale-invariant power spectrum
given by $\Delta^q_k=H^2/(4\pi^2)$~\cite{hawking}. The effects from
the quantum field $\sigma$ on the inflaton are given by the
dissipation via the kernel $G_{-}$ as well as a stochastic force
induced by the multiplicative colored noise $\xi$ with
\begin{equation}
\langle\xi(x)\xi(x')\rangle=  G_{+}(x,x'). \label{noise}
\end{equation}
Note that the white noise is uncorrelated with the colored noise 
since $\langle\xi_w(x)\xi(x')\rangle \sim 
\langle\phi(x)\sigma^2(x')\rangle=0$.

\subsection{Approximate solutions}

To solve Eq.~(\ref{lange}), let us first drop the dissipative term
which we will discuss later and consider the colored noise only.
Then, after decomposing $\phi$ into a mean field and a classical
perturbation: $\phi(\eta,\bf x)={\bar\phi}(\eta) +
{\varphi}(\eta,\bf x)$, we obtain the linearized Langevin equation,
\begin{equation}
\ddot{\varphi}+2aH\dot{\varphi}-\nabla^{2}{\varphi}+ a^{2}
m_{\varphi{\rm eff}}^2 {\varphi} = g^2 a^2 {\bar\phi}\, \xi,
\label{varphieq}
\end{equation}
where the effective mass is
$m_{\varphi{\rm eff}}^2=V''({\bar\phi})+g^{2}\langle\sigma^{2}\rangle$
and the time evolution of $\bar\phi$ is governed by $V({\bar\phi})$.
The equation of motion for $\sigma$ from which we construct its Green's
function can be read off from its quadratic terms in the
Lagrangian~(\ref{model}) as
\begin{equation}
\ddot{\sigma}+2aH\dot{\sigma}-\nabla^{2}{\sigma}+ a^{2}
m_{\sigma}^2 {\sigma} = 0.
\label{sigmaeq}
\end{equation}
Let us decompose
\begin{eqnarray}
Y(x)&=&\int\frac{d^3{\bf k}}{(2\pi)^{3\over 2}} Y_{\bf
k}(\eta)\,e^{i{\bf k}\cdot{\bf x}}, \quad
{\rm where}\;Y=\varphi,\xi, \nonumber \\
\sigma(x)&=&\int\frac{d^3{\bf k}}{(2\pi)^{3\over 2}} \left[b_{\bf
k}\sigma_k(\eta)\,e^{i{\bf k}\cdot{\bf x}} + {\rm h.c.}\right],
\end{eqnarray}
where $b_{\bf k}^\dagger$ and $b_{\bf k}$ are creation and
annihilation operators satisfying $[b_{\bf k},b_{{\bf
k}'}^{\dagger}]= \delta({\bf k}-{\bf k}')$. 
Then, the solution to Eq.~(\ref{varphieq}) is obtained as
\begin{equation}
\varphi_{\bf k} (\eta)= -ig^2\int_{\eta_i}^{\eta}
d\eta' a^4(\eta') {\bar\phi}(\eta') \xi_{\bf
k}(\eta')\left[\varphi_k^{1}(\eta')\varphi_k^{2}(\eta)
                   - \varphi_k^{2}(\eta')\varphi_k^{1}(\eta)\right],
\label{varphisol}
\end{equation}
where the homogeneous solutions $\varphi_k^{1,2}$ are given by
\begin{equation}
\varphi_k^{1,2}={1\over2a} (\pi|\eta|)^{1\over 2}
                        H_\nu^{(1),(2)}(k\eta).
\end{equation}
Here $H_\nu^{(1)}$ and $H_\nu^{(2)}$ are Hankel functions of the
first and second kinds respectively and $\nu^2=9/4-m_{\varphi{\rm
eff}}^2/H^2$. In addition, we have from Eq.~(\ref{sigmaeq}) that
\begin{equation}
\sigma_{ k} ({\eta})={1\over2a} (\pi|\eta|)^{1\over 2} \left[c_1
H_\mu^{(1)}(k\eta)+c_2 H_\mu^{(2)}(k\eta)\right],
\end{equation}
where the constants $c_1$ and $c_2$ are subject to the
normalization condition, $|c_2|^2 -|c_1|^2=1$, and
$\mu^2=9/4-m_{\sigma}^2/H^2$.

\section{Power spectrum of passive inflaton fluctuations}

Now we are able to calculate the power spectrum of the perturbation
$\varphi$. To maintain the slow-roll condition: $m_{\phi{\rm
eff}}^2=m_{\varphi{\rm eff}}^2\ll H^2$ (i.e., $\nu=3/2$), we require
that $g^2<1$ and $m_{\sigma}^2>H^2$. The latter condition
limits the growth of $\langle\sigma^{2}\rangle$ during inflation to
be less than about $10^{-2}H^2$~\cite{ford,kkk}.  This amounts to a
small contribution to the slow-roll parameter $\eta\equiv M_{Pl}^2
V''/V$, where $M_{Pl}$ is the reduced Planck mass, of order
$10^{-2}g^2/3<0.3\%$ consistent with the WMAP measurements of the
density power spectral index that determine the slow-roll parameters
only up to a few percent level~\cite{wmap}.

\begin{figure}
\leavevmode
\hbox{ \epsfxsize=4.0in
\epsffile{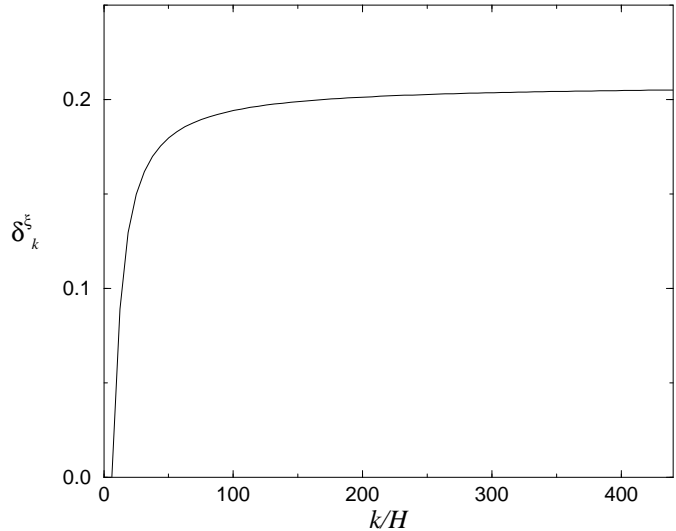}}
\caption{Power spectrum of the noise-driven inflaton fluctuations
$\delta^\xi_k\equiv 4\pi^2\Delta^\xi_k/g^4{\bar\phi}_0^2$.
The starting point, $k/H=2\pi$, corresponds to the $k$-mode that
leaves the horizon at the start of inflation.}
\label{ps}
\end{figure}

In Eq.~(\ref{sigmaeq}), we have not considered mass corrections to 
$m_{\sigma}^2$ from the mean inflaton field, $g^{2}{\bar\phi}^2$, 
and the mass renormalization due to quantum fluctuations of the inflaton,
$g^2\langle\varphi_q^{2}\rangle$.
Under the slow-roll condition, $\langle\varphi_q^{2}\rangle$ grows
linearly as $H^3t/4\pi^2$~\cite{ford,kkk} and thus
$\langle\varphi_q^{2}\rangle\simeq H^2$ after about $60$ efoldings
(i.e., $Ht\simeq 60$).  Therefore,
as long as $g^2{\bar\phi}^2 \le 2H^2$ for the period during which 
those $k$ modes of cosmologically
relevant scales cross out the horizon, we can conveniently choose
$m_{\sigma}^2=2H^2$ (i.e., $\mu=1/2$) for which $\sigma$
takes a very simple form. After then, $g^2{\bar\phi}^2$ may grow to
a value much bigger than $H^2$ and thus the effective mass of $\sigma$
becomes much larger than $H^2$. If so, this large mass will suppress
the growth of $\langle\sigma^{2}\rangle$~\cite{kkk} and may diminish
the effect of the noise term. From now on, let us consider only the
relevant period with $g^2{\bar\phi}^2 \le 2H^2$.  It was shown that
when $\mu=1/2$ one can select the Bunch-Davies vacuum (i.e., $c_2=1$
and $c_1=0$)~\cite{kkk}. Hence, using Eqs.~(\ref{noise}) and
(\ref{varphisol}), we obtain
\begin{equation}
\langle\varphi_{\bf k}(\eta)\varphi_{{\bf k}'}^*(\eta)\rangle
=\frac{2\pi^2}{k^3}\Delta^\xi_k(\eta) \delta({\bf k}-{\bf k}'),
\end{equation}
where the noise-driven power spectrum is given by
\begin{equation}
\Delta^\xi_k(\eta)=\frac{g^4 z^2}{8\pi^4} \int_{z_i}^z dz_1
\int_{z_i}^z dz_2 \, {\bar\phi}(\eta_1) {\bar\phi}(\eta_2) \frac{\,
\sin z_{-}}{z_1 z_2 z_{-}}  \left[\sin(2\Lambda
z_{-}/k)/z_{-}-1\right] F(z_1) F(z_2), \label{pseq}
\end{equation}
where $z_{-}=z_2-z_1$, $z=k\eta$, $z_i=k\eta_i=-k/H$, $\Lambda$
is the momentum cutoff introduced in the evaluation of the ultraviolet
divergent $k$-integration of $\sigma_k$ in the Green's 
function~(\ref{grfct}), and
\begin{equation}
F(y)=\left(1+\frac{1}{yz}\right)\sin(y-z)+ \left({1\over
y}-{1\over z}\right)\cos(y-z).
\end{equation}
Note that the term $\sin(2\Lambda z_{-}/k)/z_{-} \simeq
\pi\delta(z_{-})$ when $\Lambda\gg k$, so $\Delta^\xi_k(\eta)$ is
insensitive to $\Lambda$. Both
${\bar\phi}(\eta_1)$ and ${\bar\phi}(\eta_2)$ in Eq.~(\ref{pseq})
can be approximated as a constant mean field ${\bar\phi}_0$.
It is because we are concerned with large scales at
which the rate of change of the mean field at horizon-crossing 
is given by $d{\bar\phi}/d\ln k\simeq -\sqrt{2\epsilon}M_{Pl}$, 
where the slow-roll parameter $2\epsilon\equiv M_{Pl}^2 (V'/V)^2$.
The low value of $\epsilon$ is crucial in determining the
observed density power spectrum consistent with current 
measurements~\cite{lid}. In fact, it is found to be consistent
with zero up to the scale near the first CMB
Doppler peak in the WMAP measurements~\cite{wmap}.
Then, we plot $\Delta^\xi_k(\eta)$ at the
horizon-crossing time given by $z=-2\pi$ versus $k/H$ in
Fig.~\ref{ps}. The figure shows that the noise-driven fluctuations
depend on the onset time of inflation and approach asymptotically to
a scale-invariant power spectrum $\Delta^\xi_k\simeq
0.2g^4{\bar\phi}_0^2/(4\pi^2)$ at large $k$.

At this point, let us examine the dissipation term in the Langevin
equation~(\ref{lange}), which is actually divergent. We have removed
the divergency by using the regularization method~\cite{vil,kkk}
that sets the ultraviolet cutoff $\Lambda=He^{Ht}$, 
which includes all the modes with wavelength greater
than the horizon at time $t$ during inflation, as these are the ones 
responsible for the growth of $\langle\sigma^{2}\rangle$
(Note that we have used the same method to regularize the divergent
$\langle\sigma^{2}\rangle$ above). Hence, we have found that this
term only contributes a mass correction of about
$10^{-2}g^4{\bar\phi}_0^2$ to $m_{\varphi{\rm eff}}^2$ as well as a
small friction term of order $10^{-2}g^4{\bar\phi}_0^2a\dot{\phi}/H$
to Eq.~(\ref{lange}). This term also gives a correction to the
slope of the inflaton potential $V'(\phi)$ of order
$10^{-2}g^4{\bar\phi}_0^3$, which in turn changes the slow-roll
parameter $\epsilon$ by $10^{-4}g^8{\bar\phi}_0^6/(2H^4 M_{Pl}^2)$.
All of these corrections can be neglected as
long as $g^2{\bar\phi}_0^2\le 2H^2$.

\section{Effects on large-scale CMB anisotropy}

\begin{figure}
\leavevmode \hbox{
\epsfxsize=4.0in
\epsffile{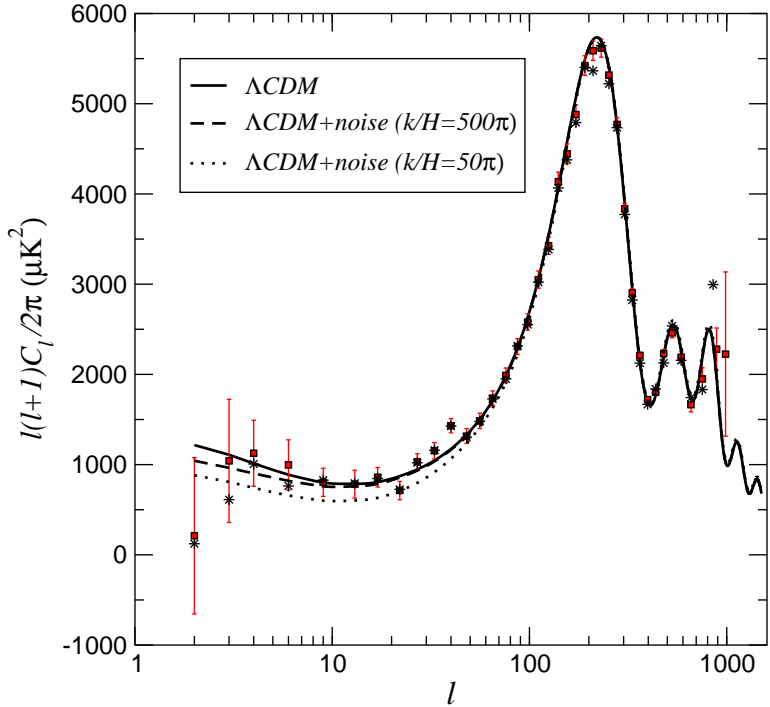}}
\caption{CMB anisotropy in the $\Lambda$CDM model with the
density power spectrum $P_k=P^q_k+P^\xi_k$.
The solid curve is the $\Lambda$CDM model with a scale-invariant
$P^q_k$ induced by quantum fluctuations. The dashed and dotted curves
represent respectively noise-driven $P^\xi_k$'s with $k/H=500\pi$
and $k/H=50\pi$ corresponding to $0.05$Mpc$^{-1}$.
We normalize all the anisotropy spectra at the first Doppler peak.
Also shown are the three-year WMAP data including error bars and
the first-year WMAP data denoted by stars~\cite{wmap}.}
\label{cl}
\end{figure}

Our results show novel effects on the large-scale density
perturbation. Within the present model, the value of
$g^4{\bar\phi}_0^2$ is not fixed except that $g^2{\bar\phi}_0^2\le
2H^2$, which is to keep the condition, $m_{\sigma}^2=2H^2$,
as well as to make sure that the time-dependent corrections to both the 
mass scales and the slow-roll parameters are small.
If $g^2\lesssim 1$ and $g^4{\bar\phi}_0^2\simeq
2H^2$, then $\Delta^\xi_k$ can be as large as $0.4\Delta^q_k$ at
large $k$. Notice that the perturbative expansion parameter is
$g^2/(2\pi)\ll 1$ such that higher-order contributions in $g$ can be
safely ignored. In the standard slow-roll inflation, the density
power spectrum induced by the active quantum fluctuations is nearly
scale-invariant and is given by $P^q_k=X({\bar\phi})\Delta^q_k$,
where $X({\bar\phi})$ is determined by the slow-roll
kinematics~\cite{olive}, whereas the noise-driven density power
spectrum, $P^\xi_k=X({\bar\phi})\Delta^\xi_k$, is blue-tilted on
large scales. Assuming a total density power spectrum
$P_k=P^q_k+P^\xi_k$ with $X({\bar\phi}=\bar{\phi}_0)$ set to
constant and using the set of cosmological parameters measured by
WMAP~\cite{wmap}, we have run the CMBFAST numerical codes~\cite{sel}
to compute the CMB anisotropy power spectrum as shown in
Fig.~\ref{cl}, where we have set $g^4{\bar\phi}_0^2=2H^2$. We can
see that a scale-invariant power spectrum at small scales can be
achieved by properly choosing the initial time of inflation. For the
dashed (dotted) curve, the Fourier mode $k/H=500\pi$ ($k/H=50\pi$)
corresponding to the physical scale of $0.05$Mpc$^{-1}$ crosses out
the horizon at about $5.5$ (3.2) efoldings. Meanwhile, the
suppressed large-scale density perturbation can account for the low
CMB low-$l$ multipoles.

\section{Implications}

Let us briefly mention other implications to the inflationary
cosmology which deserve investigations to further constrain the
free parameters here. We rewrite
$P_k=P^q_k(1+\Delta^\xi_k/\Delta^q_k)$. Therefore, the energy scale
of inflation inferred from measurements of the density power
spectrum will be lower than the standard slow-roll prediction by a
factor $(1+\Delta^\xi_k/\Delta^q_k)^{-1/4}\simeq 0.92$. The spectal
index is then given by
\begin{equation}
n(k)-1\equiv \frac{d\ln P_k}{d\ln k}= \frac{d\ln P^q_k}{d\ln k}
+ \frac{d}{d\ln k}\ln\left(1+\frac{\Delta^\xi_k}{\Delta^q_k}\right).
\label{index}
\end{equation}
The standard slow-roll inflation predicts that $|n(k)-1|\gg |d\ln
n(k)/d\ln k|$, although the three-year WMAP results have indicated
that $|n(k)-1|\simeq |d\ln n(k)/d\ln k|$~\cite{wmap}. There have
been several approaches to reduce the discrepancy by either
violating or generalizing the slow-roll condition for
$P^q_k$~\cite{leach}. Unlike these kinematic approaches, the
additional term containing $\Delta^\xi_k$ in Eq.~(\ref{index})
offers a new dynamical source for breaking the scale invariance. It
is worthwhile to study the overall effect of the corrections induced
by the dissipation, the static mass approximation, and the static
mean field approximation to the spectral index. Especially, we
expect a running spectral index at large $k$ due to dissipational
effects on the fluctuations at late times. Furthermore, it is
interesting to consider the three-point function $\langle\varphi_{\bf
k_1}\varphi_{\bf k_2}\varphi_{\bf k_3}\rangle$ by invoking
higher-order terms in the presence of inflaton mean field.

\section{Conclusions and Dsicussions}

We have discussed the effect of an interacting inflaton to the
cosmological density perturbation. The passive inflaton fluctuations
induced by the interaction are found to be blue-tilted on
large-scales. This results in a suppression of the large-scale CMB
anisotropy that may be relevant to the observed low quadrupole in
the WMAP CMB anisotropy data. Interestingly, the observed low CMB
quadrupole may open a window on the physics of the first few
efoldings of inflation.  
The results are obtained by solving the
Langevin equation~(\ref{lange}) of the inflaton arising from its
coupling to a massive quantum field with mass of the order of the
Hubble scale. Here we have not specified the inflaton potential.
However, we restrict ourselves to the situation typically for a
small-field inflation, in which the mean field ${\bar\phi}^2\simeq H^2$ 
when inflaton modes of cosmologically interesting scales cross out the
horizon. Presently, observational data prefer a slow-roll inflaton
potential; however, the exact form of the potential is still
elusive. We expect that the above mentioned situation
can be realized in certain inflationary models.
Moreover, our assumption on the values of the introduced parameters
allows us to obtain an analytic solution of the Langevin equation.
So, it would be interesting to relax this assumption by performing a
full analysis of the cosmological effects due to the interacting
quantum environment in a viable inflation model. 

In fact, one can also consider the effect of a coupling scalar field 
to the inflaton in the context of large-field inflation such as chaotic
inflation~\cite{lid}. 
We have shown that the corrections from the coupling scalar field
to the slow-roll parameters are small for $g^2{\bar\phi}_0^2\le 2H^2$.
Hence, the coupling constant $g$ can be small for those large values of 
$\phi_0$ of order of $M_{Pl}$ as usually found in chaotic inflation; 
however, this will give an insignificant correction to the density power 
spectrum. For a strong coupling of $g^2\lesssim 1$ as found for example 
in the hybrid inflation~\cite{hybrid}, the field $\sigma$ obtains a hugh 
effective mass from the inflaton mean field much greater than
$H$. This case goes beyond our assumption and needs further study.
Finally, although we have worked with a
simple inflaton-scalar interaction, our results should be generic to
any interacting model.

\begin{acknowledgments}
We are indebted to B.-L. Hu for bringing the colored noise to
our attention and many encouragements.
We also thank L. Ford and R. Rivers for useful discussions.
This work was supported in part by the National Science Council,
Taiwan, ROC under the Grants NSC 95-2112-M-001-052-MY3 (K.W.N.), 
NSC 95-2112-M-003-002 (W.L.L.), and NSC-95-2112-M-259-011-MY2 (D.S.L.).

\end{acknowledgments}

\end{document}